
\magnification=\magstep 1
\baselineskip=18 pt
\font\bigbf = cmbx10 scaled\magstep 1
 1
 1

\line{}

\noindent

\centerline{\bigbf \hfill Large N renormalization group study of \hfill}
\centerline{\bigbf \hfill the commensurate dirty boson problem \hfill}

\vskip 1.0cm

\centerline{\hfill Yong Baek Kim and Xiao-Gang Wen \hfill}
\centerline{\hfill \it Department of Physics, Massachusetts Institute of
Technology \hfill}
\centerline{\hfill \it Cambridge, Massachusetts 02139 \hfill}
\centerline{\hfill August 26, 1993 \hfill}

\vskip 1.0cm

\centerline{\hfill ABSTRACT \hfill}

\midinsert
\narrower
{\noindent\tenrm

We use a large N renormalization group (RG) method to
study a model of interacting boson system with a quenched
random potential. In the absence of impurities, the pure boson system
has a critical point that describes the superfluid-Mott-insulator
(SF-MI) transition. The SF-MI transition of $d$ dimensional bosons
belongs to the $(d+1)$ dimensional $XY$ model universality class.
In this paper, we study the dirty-boson critical points in the
neighborhood of this pure SF-MI critical point.
In general, the on-site random potential in the original lattice model
gives two types of randomness in the effective field theoretic action.
One is the randomness in the effective on-site repulsion $w({\bf x})$
and the other is the randomness of the chemical potential $u({\bf x})$.
It turns out that $d = 2$ is the critical dimension for both types of
disorder but the roles of these two types of disoder are reversed
as $d = 2$ is crossed.
Applying $\epsilon = d-2$ expansion, we found coupled RG
equations for both kinds of randomness which reveal several non-trivial
critical points.
All the weak random fixed points we found
have three or more relevant directions.
We conclude that the direct SF-MI transition is unlikely
to occur near two dimensions.

}
\endinsert

\vfill\vfill\vfill
\break

\vskip 0.5cm

\centerline{\bf I. INTRODUCTION}

The problem of repulsively interacting bosons in a random potential
has been the subject of intense research recently [1-10]. This so-called
dirty boson problem contains the essential difficulty of understanding
interplay between interaction effect and randomness. One of the reasons
why this problem is very challenging is that there is no sensible
non-interacting limit for disordered bosons.
That is, the zero interaction limit is
pathological in the sense that the bosons will condense into the
lowest localized state around a small region.
For the metal-insulator transition of electrons,
which has been understood better, disorder alone or interaction alone
can induce localization of electrons and drive the electrons to
the Anderson insulator (AI) or the Mott insultor (MI) state [11].
The interplay between these two effects has been studied during the
last decade but there are still open problems [11].

The dirty boson problem has direct experimental realization on ${}^4 He$
in Vycor glass and in other porous media [12,13]. This disordered interacting
boson problem may be used to
understand the superconductor-insulator transition in the disordered
thin films [14] and short coherence length superconductors [15].
Recently Wen and Wu [16] showed that the superfluid-Mott insulator
transition of bosons with the Chern-Simons gauge field can describe
the transition between quantum Hall (QH) states in the absence of disorder.
Therefore, the dirty boson problem with the Chern-Simons gauge field
is intimately to the quantum Hall transitions in the presence of
random impurities. The QH-MI transition for the pure system is also
studied by Chen {\it et al.} [17] in which they studied
fermions with the Chern-Simons gauge field.

Following Fisher {\it et al.} [1,2], we can write the Hamiltonian of
the interacting lattice bosons in a random on-site potential as
$$
H = H_0 + H_1 \ ,
$$
$$
\eqalign{ H_0 &= - \sum_{i} \ ( -J_0 + \mu + \delta \mu_i ) \ {\hat n}_i
\ + \ {1 \over 2} \ V \ \sum_{i} \ {\hat n}_i ( {\hat n}_i - 1 ) \ ,  \cr
H_1 &= - \sum_{i,j} \ J_{ij} \ ( b^{\dagger}_i b_j + h.c. ) \ , }
\eqno{(1)}
$$
where ${\hat n}_i = b^{\dagger}_i b_i$ and $b^{\dagger}_i$ is the boson
creation operator at the site $i$. $\mu$ is the average chemical
potential that fixes the boson density and $\delta \mu_i$ is the
random on-site potential with zero average. $J_{ij}$ is the
hopping matrix element and $J_0 = \sum_{j} J_{ij}$.
In order to study the critical phenamena of the system, it is
convenient to find the effective field theory.
We will summarize the approach of Fisher {\it et al.} [1,2].
First the off-site hoping term in $H_1$ is decoupled by
introducing the Hubbard-Stratanovich field $\Psi_i$, then
the resulting action can be expanded in terms of $\Psi_i$.
Since $\Psi_i$ is linearly related to $\langle b_i \rangle$
for small $\langle b_i \rangle$, the field $\Psi_i$ can be
identified as a superfluid (SF) order parameter.
It was shown [2] that the effective field theoretic action
is given by
$$
\eqalign{ S &= \int {d^d k \over (2 \pi)^d}
\ {d \omega \over 2 \pi} \ ( k^2 + \omega^2 )
\ |\Psi (k,\omega)|^2 \ + \ \sum_i \ ( {\bar w} + w_i ) \
|\Psi_i (\tau)|^2  \cr
&+ \ \sum_i \int d\tau \ ( {\bar u} + u_i ) \ \Psi^*_i (\tau) \
\partial_{\tau} \Psi_i (\tau)
\ + \ g \sum_i \int d\tau \ |\Psi_i (\tau)|^4 \ , }
\eqno{(2)}
$$
where $w_i, u_i$ are random functions of $i$
with zero average.

It is established [2] that SF-MI transition of the pure system  has
two uninversality classes, the commensurate case and the incommensurate
case. In the commensurate case, the superfluid density commensurates
with a periodic potential.
In this case, SF-MI transition is described by a tricritical point
which belongs to the universality class of $(d+1)$ dimensional $XY$ model
with the dynamical exponent $z = 1$.
In the incommensurate case, the SF-MI transition happens on a line
in the ${\bar w}-{\bar u}$ plane.
It is argued [2] that the generic SF-MI
transition should be the latter case ( with $z = 2$ at the transition )
rather than the former case.

One natural question is that how disorder affects these two
different SF-MI transitions. The destruction of superfluid in the
presence of disorder brought the concept of the Bose glass (BG) phase [4,18]
in which bosons are localized by disorder.
In seminal papers, Fisher and coworkers [1,2] suggested a scaling theory
of SF-BG transition. They argued that superfluid-insulator tansition should
occur through the Bose glass phase and the generic transition should be
described by the action (2) with ${\bar u} + u_i \not= 0$ which does
not have space-time isotropy.
In the scaling theory, they postulated that the compressibility
is totally due to the phonon mode and one of the main result of this
assumption is that the dynamical exponent $z=d$. The simulation of the
quantum rotor model [3] and some renormalization
group calculation [7] partially supported this picture although
a recent quantum Monte-Carlo calculation [9] contradicts to
these results.
The earlier work of Ma, Halperin and Lee (MHL) [4] was reexamined and
the importance of the term that is linear in $\omega$ is emphasized.
However, it was also argued that MHL theory may apply to the possible
direct SF-MI transition at the commensurate case
( particle-hole symmetric case ).

We can see that the dirty boson effective action has the strict
particle-hole symmetry if ${\bar u} + u_i = 0$. It was also
argued that the general commensurate case corresponds to the weaker
particle-hole symmetric case [2],
{\it i.e.}, ${\bar u} = 0$ but $u_i \not= 0$.
Are the transitions of particle-hole symmetric and asymmetric cases in
the same universality class ?  Originally Fisher and coworkers [1,2]
prefered that even arbitrarily weak disorder will induce the Bose glass
phase for both of the incommensurate and commensurate cases (Fig.1 (a)).
However, numerical calculations in [8] have not revealed the intervening
Bose glass phase in the commensurate case, although the
superfliud-insulator transition indeed occur through the
Bose glass phase in the incommensurate case (Fig.1 (b)).
Singh and Rokhsar [5]
performed a real space renormalization group analysis for the
commensurate case and they found the direct transition from SF to MI
when the disorder is sufficiently weak ; the Bose glass is found beyond
a threshold ( Fig.1 (a) or Fig.1 (b) depending on the impurity
strength).
Zhang and Ma [6] considered hard core bosons with disorder. In this
real space renormalization group analysis of a quantum spin-${1 \over 2}$
$XY$ model with transverse random field ( the hard core boson model is
mapped to this model), they concluded that commensurate and incommensurate
cases are in the same universality class and the SF-insulator transition
occurs alwalys from the Bose glass phase (Fig.1 (a)).
The conventionl renormalization group calculation by Weichman and Kim [7]
in which they used double dimensional expansion around $d=4$ partially
supported the original picture [1,2] for the general dirty boson problem
although some technical problems exist.

In this paper, we are going to study a large N generalization of
the original action.
By doing ${1 \over N}$ expansion, we can treat the
interaction non-pertubatively in the coupling constant. Both types
of disorder are assumed to be weak and we do the pertubation in the
strength of two types of disorder.
The large N generalized action of the original dirty boson
model in the Eucledian space is given by
$$
\eqalign{
S &= \ \int {d^d k \over (2\pi)^d} \ {d\omega \over 2\pi} \
( \omega^2 + v_0^2 k^2 + {\bar w} ) \ |\phi_i (k,\omega)|^2 \  \cr
&+ \ \int d^d x  \ d\tau \ \left [ \ ({\bar u} + u({\bf x}))
\ \phi_i^{\dagger} \partial_{\tau} \phi_i \
+ w({\bf x}) \ \phi_i^{\dagger} \phi_i \ + \
{g_0 \over N} \ ( \phi_i^{\dagger} \phi_i )^2 \ \right ] \ , }
\eqno{(3)}
$$
where $i = 1, \cdots, N$ and $N = 1$ corresponds to the original model.
$u({\bf x})$ and $w({\bf x})$ are gaussian random functions of {\bf x}
with zero mean and their variances are given by
$\langle u({\bf x}) u({\bf y}) \rangle =
U_0 \ \delta^d ( {\bf x} - {\bf y} )$ and
$\langle w({\bf x}) w({\bf y}) \rangle =
W_0 \ \delta^d ( {\bf x} - {\bf y} )$ respectively,
where $\langle \cdots \rangle$ means the random average.

When ${\bar u} + u({\bf x}) = {\bar w} + w({\bf x}) = 0$, (3) describes the
muticritical point of SF-MI transition of the pure boson system.
Near $d=2$, both of ${\bar u}$ and ${\bar w}$ terms are strongly relevant.
In this paper, we will take ${\bar u} = {\bar w} = 0$ and study the
RG flow of $U_0$, $W_0$ and $v_0$. The physical meaning of setting
${\bar u} = 0$ is the following. We tune the chemical potential $\mu$ to
make the average boson density alwalys commensurate with the lattice.
Thus, we will call (3) with ${\bar u} = 0$ the commensurate dirty boson
theory.

Our RG calculations are done at the critical point and
with the renormalized mass term ${\bar w} = 0$
in the course of the renormalization.
Since the effect of $\phi^4$ term is calculated exactly at each order
of $1/N$, the coupling constant $g_0$ is not renormalized.
Following Ref.[19] and using dimensional regularization,
we move an infrared scale $\mu$ in the renormalized theory
with fixed bare parameters to obtain the RG flow of the renormalized
parameters.
Introducing dimensionless measures of disorder,
${\tilde U} = { U \over v^2 } \mu^{d - 2}$ and
${\tilde W} = { W v^2 \over g_0^2 } \mu^{2 - d }$, we
performed $\epsilon = d - 2 $ expansion.
The resulting renormalization group equation up to
$({1 \over N})^0$ order is found to be
$$
\eqalign{ {d {\tilde W} \over d l} \ &=  \ \epsilon {\tilde W} +
a {\tilde U} {\tilde W} + b {\tilde W}^2 \ ,
\cr
{d {\tilde U} \over d l} \ &= \ -  \epsilon {\tilde U} +
b {\tilde W} {\tilde U} -  a {\tilde U}^2  \ , \cr
{d \ ({\rm ln} \ v) \over d l} \ &= \ -  b {\tilde U} +  c {\tilde W} \ , }
\eqno{(4)}
$$
where $a={1 \over 2\pi}, b={128 \over \pi}, c={64 \over \pi}$
and $l$ is the logarithmic measure of the RG flow.
This is the central result of this paper.

Looking at $\epsilon \ge 0$ case, we can immediately see that there is only
the pure fixed point which is given by ${\tilde U}^*=0, {\tilde W}^*=0$.
At this trivial fixed point, ${\tilde U}$ is irrelevant
and ${\tilde W}$ is relevant so that the
RG flow goes to the strong randomness regime where our RG breaks
down. However, for $\epsilon < 0$, there are three fixed points.
All of them have at least one relevant direction in the
${\tilde U}-{\tilde W}$ plane. Thus, including ${\bar u}$ and ${\bar w}$,
these fixed points in the original theory (3) has at least
three relevant directions.
Therefore, in both cases, the direct SF-MI transition in the
${\tilde W}-{\tilde U}$ plane
is unlikely to occur due to the absence of the weak random fixed point
with two or less relevant directions.
The superfluid insulator transition
is alwalys governed by a strong random fixed point which cannot be
reached by weak randomness expansion.
More details of the RG flow will be discussed later.

The organization of this paper is as follows. In section II, we
consider a restricted model in which $u({\bf x})=0$ and
show the basic formalism we used. Here, we also examine some possible
effects of long range interactions.
The RG calculation for the generic commensurate dirty
boson problem up to $(1/N)^0$ order is presented and
the RG equation is calculated in section III. we also discuss
about the results.
In section IV, $1/N$ correction due to $\phi^4$ interaction
is considered.
In section V, we summarize and conclude this paper.

\vskip 0.5cm

\centerline{\bf II. RENORMALIZATION GROUP ANALYSIS FOR}
\centerline{\bf THE STRONG PARTICLE-HOLE SYMMETRIC MODEL }

In this section, we consider a rather restricted model in which we set
$u({\bf x})=0$. This corresponds to the {\it strong} particle-hole
symmetric case in the sense that more general model requires the
particle-hole symmetry only in the average sense, {\it i.e.},
$\langle u({\bf x}) \rangle = 0$ and allows the local breaking of
the symmetry at each site.
We would like to consider more general long range interactions
$\int \ d^d x \ d^d y \ \phi^2 ({\bf x}) \ V({\bf x}-{\bf y})
\ \phi^2 ({\bf y})$
with $V({\bf x}) \propto g_0 / |{\bf x}|^{d - \lambda}$ and
$V(q) = g_0 / q^{\lambda}$ ( $0 \le \lambda < 1/2$ ).
The $\lambda = 0$ case corresponds to the usual short range interaction.
However, it should be mentioned that these are not the true long range
interactions of bosons because the true one should be $V({\bf x}-{\bf x}')
\rho ({\bf x}) \rho ({\bf x}')$ and $\rho ({\bf x}) = i ( \phi^{\dagger}
\partial_0 \phi - \partial_0 \phi^{\dagger} \phi )$.
But the long range interaction between densities will induce the long
range interaction in the $\phi^4$ term.

In order to perform the RG calculation in the critical theory, we need
to introduce an arbitrary infrared mass parameter $\mu$ [19].
We will use the dimensional regularization method to calculate relevant
divergent diagrams. The renormalization of the theory is given
by the renormalization of two point $\Gamma^{(2)}$ and
four point $\Gamma^{(4)}$ vertices. We will take the inverse of the
full propagator as a two point vertex and the two-boson scattering
amplitude as a four point vertex function.
The relation between the bare theory and the renormalized theory
is given by
$$
\Gamma^{(N)}_{\rm bare}(q,\omega;\Lambda) \ = \
Z^{-N/2}(\Lambda / \mu) \ \Gamma^{(N)}(q,\omega;\mu) \ ,
\eqno{(5)}
$$
where $\Gamma^{(N)}_{\rm bare}$ and $\Gamma^{(N)}$ represent the bare
and the renormalized vertices respectively.
We found that appropriate renormalization condition for
$\Gamma^{(2)}$ can be chosen as
$$
\eqalign{ {\partial \over \partial \omega^2}
\Gamma^{(2)}{\Bigr |}_{q = \mu, \ \lim_{\alpha \rightarrow 0}
\omega = \alpha v \mu} \ &= \ 1  \ , \cr
{\partial \over \partial q^2}
\Gamma^{(2)}{\Bigr |}_{q = \mu, \ \lim_{\alpha \rightarrow 0}
\omega = \alpha v \mu} \ &= \ v^2 \ . }
\eqno{(6)}
$$
Renormalization condition for the scattering amplitude will be discussed
later.
Also, following standard procedure, we require the independence of the
bare theory with respect to $\mu$,
$$
\mu {d \over d \mu}{\Bigr |}_{\Lambda} \Gamma^{(N)}_{\rm bare} = 0 \ ,
\eqno{(7)}
$$
where $\Lambda$ is the mass parameter of the bare theory.

Let us start with the evaluation of the self energy to the $(1/N)^0$ order.
The $(1/N)^0$ order self energy diagram is given by Fig.2 (a).
The polarization bubble $\Pi_1 (q,\omega=0)$ in Fig.2 (b) is calculated as
$$
\eqalign{
\Pi_1 (q,\omega=0) &= \int \ {d^d k \over (2 \pi)^d} \ {d\nu \over 2 \pi}
\ {1 \over \nu^2 + v_0^2 k^2} \ {1 \over \nu^2 + v_0^2 (k-q)^2} \cr
&= {c_1 \over v_0^3} q^{d-3} \ ,
}
$$
$$
c_1 = {1 \over (4 \pi)^{{d+1 \over 2}}} { \Gamma ({3-d \over 2})
\Gamma^2 ({d-1 \over 2}) \over \Gamma (d-1)  } \ .
\eqno{(8)}
$$
Assuming $3 + \lambda > d$, the diagram in Fig.2 (c) can be approximated as
$$
W_0 \left ({1 \over 1 + \Pi_1 (q,\omega=0) (g_0 / q^{\lambda}) } \right )^2
\ \approx \ W_0 {v^6_0 \over c_1^2 g_0^2} q^{2 ( 3 + \lambda - d)} \ .
\eqno{(9)}
$$
Using this result, we can evaluate the self energy as
$$ \eqalign{ \Sigma_1 (q,\omega) &= \int {d^d k \over (2 \pi)^d}
\ {1 \over \omega^2 + v_0^2 k^2}
\left ( {W_0 v_0^6 \over c_1^2 g_0^2} \right )
(k - q)^{2 ( 3 + \lambda - d)} \cr
&\approx \left ( {W_0 v_0^6 \over c_1^2 g_0^2} \right )
( c_2 q^{4 + 2\lambda - d} + c_3 {\omega^2 \over v^2_0}
q^{2 + 2\lambda - d} ) \ , }
$$
$$
\eqalign{ c_2 &= {1 \over (4 \pi)^{d/2}} \ {\Gamma (d/2 - 2 - \lambda)
\ \Gamma (3 + \lambda - d/2) \ \Gamma (d/2 - 1) \over
\Gamma (d - 3 - \lambda) \ \Gamma (2 + \lambda) }  \ , \cr
c_3 &= {2 + \lambda - d/2 \over (4 \pi)^{d/2}} \ {\Gamma (d/2 - 2 - \lambda)
\ \Gamma (3 + \lambda - d/2) \ \Gamma (d/2 - 2) \over
\Gamma (d - 3 - \lambda) \ \Gamma (1 + \lambda)} \ , }
\eqno{(10)}
$$
where $v_0 q \gg \omega$ is assumed.
The bare two point vertex $\Gamma^{(2)}_{\rm bare}$ up to
$({1 \over N})^0$ order is
$$
\eqalign{ \Gamma^{(2)}_{\rm bare} &=
\omega^2 + v_0^2 q^2 - \Sigma_1 (q,\omega) \cr
&= \omega^2 ( 1 - {W_0 v_0^2 \over g_0^2} {c_3 \over c_1^2}
q^{2+2\lambda-d})  \cr
&+ v_0^2 q^2 ( 1 - {W_0 v_0^2 \over g_0^2} {c_2 \over c_1^2}
q^{2+2\lambda-d}) \ . }
\eqno{(11)}
$$
Note that $c_2$ and $c_3$ diverge at $d=2+2\lambda$. Therefore,
let us try $\epsilon = d-2-2\lambda$ expansion in order to handle
these divergences. Also, for convenience, let us introduce
dimensionless measures of the disorder ${\tilde W_0} =
{W_0 v_0^2 \over g_0^2} \Lambda^{2+2\lambda-d}$ in the bare theory
and ${\tilde W} = {W v^2 \over g_0^2} \mu^{2+2\lambda-d}$ in the
renormalized theory.
Adding appropriate counter terms to cancel
the ${1 \over \epsilon}$ divergences from $c_2$ and $c_3$ and
using the renormalization conditions (5) (6),
we get the following equations
$$
\eqalign{ Z &\approx 1 + {\tilde W} \ {{\tilde c_3} \over c_1^2} \ {\rm ln}
\ {\Lambda \over \mu} \ , \cr
v_0^2 &\approx Z^{-1} v^2 \left ( 1 - {\tilde W} \
{{\tilde c_2} \over c_1^2} \ {\rm ln} \
{\Lambda \over \mu} \right )^{-1} \ ,}
\eqno{(12)}
$$
where ${\tilde c_2} = \epsilon c_2$ and ${\tilde c_3} = \epsilon c_3$.
{}From $\mu$ independence of the bare parameters, we get
$$
\eqalign{ \mu {\partial \over \partial \mu} ( {\rm ln} Z )
&\approx - {\tilde W} {{\tilde c_3} \over c_1^2}  \ , \cr
\mu {\partial \over \partial \mu} v &\equiv \beta (v)
\approx {1 \over 2} \left [ {{\tilde c_2} \over c_1^2}
- {{\tilde c_3} \over c_1^2}
\right ] {\tilde W} v \ .}
\eqno{(13)}
$$

Now we are going to renormalize the four point function $\Gamma^{(4)}$.
We will take the scattering amplitude of two bosons at
$q_1=q_4=q, q_2=q_3=0,
\omega_1=\omega_2=\omega_3=\omega_4=\omega$ as our $\Gamma^{(4)}$.
First of all, let us calculate $\Gamma^{(4)}_{\rm bare}$ in the bare
theory. $\Gamma^{(4)}_{\rm bare}$ in the $({1 \over N})^0$ order
is the sum of $\Gamma^{(4)}_{{\rm bare},1}, \Gamma^{(4)}_{{\rm bare},2},
\Gamma^{(4)}_{{\rm bare},3},\Gamma^{(4)}_{{\rm bare},4}$ which are shown in
Fig.3 (a), (b), (c), (d) respectively.
Let us introduce a renormalized polarization buble ${\tilde \Pi}_1$,
a new polarization bubble $\Pi_2$ and a vertex $V_1$ which
are given by the diagrams in Fig.3 (e), (f) and (g).
$\Gamma^{(4)}_{\rm bare}$ can be calculated symbolically as
$$\eqalign{\Gamma^{(4)}_{\rm bare}
&= \Gamma^{(4)}_{{\rm bare},1} + \Gamma^{(4)}_{{\rm bare},2} +
\Gamma^{(4)}_{{\rm bare},3} + \Gamma^{(4)}_{{\rm bare},4} \cr
&\approx \left ( {1 \over 1 + g_0 ( {\tilde \Pi}_1 + \Pi_2 ) }
\right )^2 W_0 +
2 V_1 \left ( {1 \over 1 + g_0 {\tilde \Pi}_1 } \right )^2 W_0 \ .}
\eqno{(14)}
$$
The bubble ${\tilde \Pi}_1$ which is renormalized by the self energy
correction is given by
$$
\eqalign{
{\tilde \Pi}_1 (q,\omega=0)
&= \int \ {d^d k \over (2 \pi)^d} \ {d\nu \over 2 \pi}
\ {Z \over \nu^2 + v^2 k^2} \ {Z \over \nu^2 + v^2 (k-q)^2} \cr
&= Z^2 {c_1 \over v^3} q^{d-3} \ .
}
\eqno{(15)}
$$
Evaluation of $\Pi_2 (q, \omega=0)$ is a little bit long task and
the result is
$$
\eqalign{\Pi_2 (q, \omega=0) &= {W_0 v_0^6 \over g_0^2 c_1^2}
\int {d^d k \over (2\pi)^d} \ {d^d p \over (2\pi)^d} \ {d \nu \over 2\pi}
\ {1 \over v_0^2 (k-q)^2 + \nu^2} \ {1 \over v_0^2 k^2 + \nu^2} \cr
&\times {1 \over v_0^2 (k-p)^2 + \nu^2} \ {1 \over v_0^2 (k-p-q)^2 + \nu^2}
\ {1 \over p^{2(d-3-\lambda)}} \cr
&= {W_0 \over v_0 g_0^2} {e \over c_1^2}
q^{d-3} q^{2+2\lambda-d} \ ,
}
\eqno{(16)}
$$
where
$$
\eqalign{
e &= {1 \over (4 \pi)^{d + 1/2}} \ {\Gamma (1/2 - \lambda) \over
\Gamma (d-3-\lambda)} \cr
&\times
\int_0^1 dx_1 \int_0^{1-x_1} dx_2 \ x_2^{d/2-2-\lambda}
\ (1-x_2)^{2-\lambda-d/2} \cr
&\times
\int_0^1 dy_1 \int_0^{1-y_1} dy_2 \ y_2^{d/2-2-\lambda}
\ (x_2 + y_2 (1-x_2))^{-1/2} \cr
&\times \left [ x_1 (1-x_1) (1-x_2) y_2 - x_1^2 y_2^2 x_2
- 2 y_1 y_2 x_1 x_2 (1-x_2) + y_1 (1-y_1) x_2 (1-x_2)^2
\right ]^{\lambda-1/2} \ .
}
\eqno{(17)}
$$
We found that $\lambda \not= 0$ and $\lambda = 0$ cases should be treated
separately. First, for $\lambda = 0$ case, it is found that the constant
$e$ does not diverge as $\epsilon \rightarrow 0$ so that
it can be dropped in the final RG equations.
Now, let us look at the case of $\lambda \not= 0$. The strategy is that
we investigate the most divergent contributions in various limits and
add up all of the contributions.
Since the most divergent contribution comes from $x_2 \rightarrow 0$ or
$y_2 \rightarrow 0$ limit in Eq.(17) and we want just this contribution,
we can set $x_2=0$ or $y_2=0$ inside the square bracket in Eq.(17).
We found that both ways give the same answer.
Here we will take $x_2=0$ limit inside the square bracket and
multiply by $2$.
The equation (17) becomes
$$
\eqalign{
e &= {2 \over (4 \pi)^{d + 1/2}} \ {\Gamma (1/2 - \lambda) \over
\Gamma (d-3-\lambda)} \cr
&\times
\int_0^1 dx_1 \ x_1^{\lambda-1/2} (1-x_1)^{\lambda-1/2} \
\int_0^{1-x_1} dx_2 \ x_2^{d/2-2-\lambda} (1-x_2)^{2-\lambda-d/2} \cr
&\times
\int_0^1 dy_1 \ \int_0^{1-y_1} dy_2 \ y_2^{d/2-3} \cr
&\approx {2 \over (4 \pi)^{d + 1/2}} \ {\Gamma (1/2 - \lambda) \over
\Gamma (d-3-\lambda) \ (d/2 - 1 - \lambda) \ (d/2 - 2)} \cr
&\times
\int_0^1 dx_1 \ x_1^{\lambda-1/2} (1-x_1)^{d/2-3/2} \cr
&\times
\int_0^1 dy_1 \ (1-y_1)^{d/2-2} \ ,
}
\eqno{(18)}
$$
where only the leading divergence is taken in the second equation.
The evaluation of the remaining integrals
is straightforward and the result is
$$
\eqalign{
e &=
{2 \ \Gamma (1/2-\lambda) \ \Gamma (d/2-1/2) \ \Gamma (\lambda+1/2)
\ \Gamma (d/2-1)
\over (4\pi)^{d+1/2} \ (d/2-1-\lambda) \ (d/2-2) \
\Gamma (d-3-\lambda) \ \Gamma (d/2) \
\Gamma (\lambda+d/2) } \cr
&= {2 \over \epsilon} \ {2 \over (4\pi)^{d+1/2}} \
{\Gamma (1/2-\lambda) \ \Gamma (1/2+\lambda+\epsilon/2)
\ \Gamma (\lambda+1/2) \ \Gamma (-1+\lambda+\epsilon/2)
\over \Gamma (-1+\lambda+\epsilon) \ \Gamma (1+\lambda+\epsilon/2) \
\Gamma (1+2\lambda+\epsilon/2) } \ .
}
\eqno{(19)}
$$
$V_1 (q_1=q_3=q,q_2=0,\omega_1=\omega_3=\omega, \omega_2=0)$ can be evaluated
similarly as
$$
\eqalign{ V_1
&= {W_0 v_0^6 \over g_0^2 c_1^2}
\int {d^d p \over (2\pi)^d} \ {1 \over v_0^2 (q-p)^2 + \omega^2} \
{1 \over v_0^2 p^2 + \omega^2 } \
{1 \over p^{2(d-3-\lambda)}} \cr
&\approx
{W_0 v_0^2 \over g_0^2}
q^{2+2\lambda-d}
\left ( {d_1 \over c_1^2} + \left ( {\omega^2 \over v_0^2 q^2} \right )
{d_2 \over c_1^2} \right ) \ ,
}
\eqno{(20)}
$$
$$
\eqalign{
d_1
&= {1 \over (4\pi)^{d/2}} \
{\Gamma (d/2-1-\lambda/2) \ \Gamma (2-d/2+\lambda) \ \Gamma (d/2-1)
\over \Gamma (d-2-\lambda) \ \Gamma (1+\lambda) } \ , \cr
&= {1 \over (4\pi)^{d/2}} \
{\Gamma (\epsilon/2) \ \Gamma (1-\epsilon/2) \ \Gamma (\lambda+\epsilon/2)
\over \Gamma (\lambda+\epsilon) \ \Gamma (1+\lambda) } \cr
d_2 & = d_1 \lambda \left [ {2 (\lambda-1) \over d-4}
- {d-3-\lambda \over d-2-\lambda} \right ] \ ,
}
\eqno{(21)}
$$
where $v_0 q \gg \omega$ is assumed.
Note that $d_2=0$ for the short range interaction $( \lambda=0 )$.

Putting together all of these results, we can get the following
$\Gamma^{(4)}_{\rm bare} (q_1=q_4=q, q_2=q_3=0,
\omega_1=\omega_2=\omega_3=\omega_4=\omega)$,
$$
\eqalign{
\Gamma^{(4)}_{\rm bare}
&\approx \left ( {1 \over g_0 ( {\tilde \Pi}_1 + \Pi_2 )} \right )^2 W_0 +
2 V_1 \left ( {1 \over g_0 {\tilde \Pi}_1} \right )^2 W_0  \cr
&\approx {W_0 v^6 \over g_0^2 c_1^2} q^{2(3-d)} Z^{-4} \left [
\left ( 1 + {W_0 \over v_0 g_0^2} {e \over c_1^3} {v^3 \over Z^2}
q^{2+2\lambda-d} \right )^{-2} \right. \cr
&+ \left. 2 {W_0 v_0^2 \over g_0^2}
q^{2+2\lambda-d}
\left ( {d_1 \over c_1^2} + \left ( {\omega^2 \over v_0^2 q^2} \right )
{d_2 \over c_1^2} \right ) \right ] \ .
}
\eqno{(22)}
$$
The renormalization condition for $\Gamma^{(4)}$ is taken as
$$
\Gamma^{(4)} (q_1=q_4=\mu, \ q_2=q_3=0, \ \lim_{\alpha \rightarrow 0}
\omega_1=\omega_2=\omega_3=\omega_4=\alpha v \mu)
= {W v^6 \over g_0^2 c_1^2}\mu^{2(3-d)} \ .
\eqno{(23)}
$$
We introduce again ${\tilde W} = {W v^2 \over g_0^2} \mu^{2+2\lambda-d}$ and
${\tilde W_0} = {W_0 v_0^2 \over g_0^2} \Lambda^{2+2\lambda-d}$.
If appropriate counter terms were added to cancel the $1 / \epsilon$
divergences, the following equation can be obtained
$$
\eqalign{
{\tilde W_0}
&\approx {\tilde W} Z^2 \left ( {v_0 \over v} \right )^2
\left ( {\mu \over \Lambda} \right )^{d-2-2\lambda}
\left [ \ 1 + 2 {\tilde W} \ {{\tilde e} \over c_1^3} \ {\rm ln} \
{\Lambda \over \mu} -
2 {\tilde W} \ {{\tilde d_1} \over c_1^2} \ {\rm ln} \
{\Lambda \over \mu} \ \right ] \ \cr
&\approx {\tilde W}
\left ( {\mu \over \Lambda} \right )^{d-2-2\lambda}
\left [ \ 1 + {{\tilde W} \over c_1^2}
\ ( {\tilde c_3} + {\tilde c_2} + 2 {{\tilde e} \over c_1}
- 2 {\tilde d_1} ) \ {\rm ln} \ {\Lambda \over \mu}
\ \right ] \ ,
}
\eqno{(24)}
$$
where ${\tilde e} = \epsilon e$ and ${\tilde d_1} = \epsilon d_1$.
Using the fact that the bare parameters are fixed, as we change $\mu$,
we get the following equation for ${\tilde W}$
$$
\mu {\partial \over \partial \mu} {\tilde W} \equiv
\beta ({\tilde W}) = - \epsilon {\tilde W}
+ {1 \over c_1^2}
\left [ {\tilde c_2} + {\tilde c_3} +
2 {{\tilde e} \over c_1} - 2 {\tilde d_1}
\right ]  {\tilde W}^2 \ .
\eqno{(25)}
$$
Let $b$ be the reduction factor for the momentum scale from $\mu$ to
$\mu / b$, then the RG equations for ${\tilde W}$ and $v$ up to
$({1 \over N})^0$ order is
$$
\eqalign{ {d {\tilde W} \over d l} &= - \beta ({\tilde W})
= \epsilon {\tilde W} - {1 \over c_1^2}
\left [ {\tilde c_2} + {\tilde c_3} +
2 {{\tilde e} \over c_1} - 2 {\tilde d_1}
\right ]  {\tilde W}^2  \ , \cr
{d v \over d l} &= - \beta (v) = {1 \over 2 c_1^2} \left ( {\tilde c_3} -
 {\tilde c_2} \right )
v {\tilde W} \ ,}
\eqno{(26)}
$$
where $l = {\rm ln} b$ is the logarithmic measure of the RG flow.
${\tilde c_2}$ and ${\tilde c_3}$ are given by
$$
\eqalign{ {\tilde c_2} &= - {2 \over (4 \pi)^{d/2}}
\ {\Gamma (2 - \epsilon / 2) \ \Gamma ( \lambda + \epsilon / 2) \over
\Gamma (-1 + \lambda + \epsilon) \ \Gamma (2 + \lambda) } \ , \cr
{\tilde c_3} &= - {2 - \epsilon \over (4 \pi)^{d/2}}
\ {\Gamma (2 - \epsilon / 2) \ \Gamma ( -1 + \lambda + \epsilon / 2) \over
\Gamma (-1 + \lambda + \epsilon) \ \Gamma (1 + \lambda)} \ . }
\eqno{(27)}
$$

Now, the RG equations for $0 < \lambda < 1/2$ up to
$({1 \over N})^0$ order is
$$
\eqalign{ {d {\tilde W} \over d l} &= - \beta ({\tilde W})
= \epsilon {\tilde W}
- {4 \over c_1^2 \ (4 \pi)^{d/2} \ \Gamma (2+\lambda)} {\tilde W}^2 \ , \cr
{d v \over d l} &= - \beta (v) = - {1 \over c_1^2}
{2 \over (4 \pi)^{d/2}} {1 \over \Gamma (2+\lambda)}
v {\tilde W} \ ,}
\eqno{(28)}
$$
where the following coefficients are used:
$$
\eqalign{
{\tilde c_2} &= {2 \over (4 \pi)^{d/2}}
\ {1-\lambda \over \Gamma (2 + \lambda) } \ , \cr
{\tilde c_3} &= - {2 \over (4 \pi)^{d/2}}
\ {1 \over \Gamma (1 + \lambda)} \ , \cr
{\tilde d_1} &= - {\tilde c_3} \ , \cr
{{\tilde e} \over c_1} &= {4 \over (4 \pi)^{d/2}}
\ {1 \over \Gamma (1 + \lambda)} \ .
}
\eqno{(29)}
$$
For the usual short range interactions $\lambda = 0$,
$$
\eqalign{ {d {\tilde W} \over d l} &= - \beta ({\tilde W}) =
\epsilon {\tilde W} + {128 \over \pi} {\tilde W}^2  \ , \cr
{d v \over d l} &= - \beta (v) =  - {64 \over \pi} v {\tilde W} \ ,}
\eqno{(30)}
$$
where
$$
c_1 = 1/8, \ {\tilde c_2} = 1 / \pi, \ {\tilde c_3} = - 1 / \pi,
\ {\tilde d_1} = 1 / \pi
\eqno{(31)}
$$
were used. ${\tilde e} = \epsilon e$ is the order of $\epsilon$ for the
short range interaction and is dropped in the RG equation.

The RG equations (28) for $\lambda \not= 0$ tell us that the disorder is
relevant for $d > d_c = 2 + 2\lambda$, irrelevant for $d < d_c$
and marginally irrelevant for $d=d_c$.
For $d_c < d < 3 + \lambda$,
the pure fixed point ${\tilde W}^* =0$ is unstable.
However, there is a stable fixed point which is given by
${\tilde W}^* = \epsilon / Q$ where
$Q = {4 \over c_1^2 \ (4 \pi)^{d/2} \ \Gamma (2+\lambda)}$.
{}From $\beta (v)$, we can read off the dynamical exponent
$z = 1 +  {1 \over c_1^2} {2 \over (4 \pi)^{d/2}}
{1 \over \Gamma (2+\lambda)}{\tilde W}^*$.
More specifically, $z = 1 + \epsilon / 2$ at this stable fixed point.
For $d \le d_c$, the disorder is irrelevant and the pure fixed point
is stable. Therefore, we can expect the direct superfluid-Mott
insulator transition for $d \le d_c$.

Now let us look at the short range interaction case $\lambda = 0$.
{}From Eq. (30), we can see that the disorder is relevant for $d \ge 2$
and irrelevant for $d < 2$. Therefore, there is only the unstable
pure fixed point and the RG flow goes to the
strong disorder regime for $d = 2$ and slightly larger than two.
For $d < 2$, the pure fixed
point becomes stable. There is also an unstable fixed point which is
given by ${\tilde W}^* = |\epsilon| \pi / 128$. The dynamical
exponent at the unstable fixed point is $z = 1 + |\epsilon| / 2$.

\vskip 0.5cm

\centerline{\bf III. RENORMALIZATION GROUP ANALYSIS FOR}
\centerline{\bf THE COMMENSURATE DIRTY BOSON MODEL}

Now we are going to study the generic model (3) with
${\bar u} = {\bar w} = 0$.
Here, we consider the usual short range interaction $\lambda = 0$.
The disorder characterized by $u({\bf x})$ in Eq.(3) will be considered
in addition to the $w({\bf x})$ type disorder. This means that we have
to consider more diagrams that are generated by this new disorder.
The additional self energy correction due to $u({\bf x})$ type disorder is
given by the diagram in Fig.4
$$ \eqalign{ \Sigma_2 (q,\omega) &= - U_0 \omega^2
\int {d^d k \over (2 \pi)^d}
\ {1 \over \omega^2 + v_0^2 (q-k)^2} \cr
&= {U_0 \over v_0^d} c_4 \omega^2 \omega^{d-2} \ , \cr
c_4 &= - {\Gamma (1 - d/2) \over (4\pi)^{d/2}} \ , }
\eqno{(32)}
$$
The new bare two point vertex $\Gamma^{(2)}_{\rm bare}$ up to
$({1 \over N})^0$ order is
$$
\eqalign{ \Gamma^{(2)}_{\rm bare} &=
\omega^2 + v_0^2 q^2 - \Sigma_1 (q,\omega) - \Sigma_2 (q,\omega) \cr
&= \omega^2 \left [ 1 - {W_0 v_0^2 \over g_0^2} {c_3 \over c_1^2}
q^{2-d} - {U_0 \over v_0^2} c_4 \left ( \omega \over v_0
\right )^{d-2} \right ]  \cr
&+ v_0^2 q^2 \left ( 1 - {W_0 v_0^2 \over g_0^2} {c_2 \over c_1^2}
q^{2-d} \right ) \ . }
\eqno{(33)}
$$
It is clear that we need to do $\epsilon = d-2$ expansion in order to
handle the divergences. Let us introduce additional dimensionless measures
of the disorder ${\tilde U_0} = {U_0 \over v_0^2} \Lambda^{d-2}$ and
${\tilde U} = {U \over v^2} \mu^{d-2}$ in the bare and the renormalized theory.
Adding appropriate counter terms and using (5) (6) (7), we can again obtain
the following equations for $Z$ and $v$
$$
\eqalign{ \mu {\partial \over \partial \mu} ( {\rm ln} Z )
&\approx - {\tilde W} {{\tilde c_3} \over c_1^2} + {\tilde c_4}
{\tilde U} \ , \cr
\mu {\partial \over \partial \mu} v &\equiv \beta (v)
\approx {1 \over 2} \left [ {{\tilde c_2} \over c_1^2}
- {{\tilde c_3} \over c_1^2}
\right ] {\tilde W} v + {\tilde c_4} {\tilde U} v \ .}
\eqno{(34)}
$$

There can be $5$ additional diagrams that contribute to
$\Gamma^{(4)}_{\rm bare}$. Let us identify $\Gamma^{(4)}_{\rm bare,5}$,
$\Gamma^{(4)}_{\rm bare,6}$, $\Gamma^{(4)}_{\rm bare,7}$,
$\Gamma^{(4)}_{\rm bare,8}$, $\Gamma^{(4)}_{\rm bare,9}$ as the diagrams in
Fig.5 (a), (b), (c), (d), (e) respectively. First of all, let us
evaluate the new vertex $V_2$ of Fig.6 (a)
$$
\eqalign{ V_2
&= - U_0 \omega^2
\int {d^d p \over (2\pi)^d} \ {1 \over v_0^2 (q-p)^2 + \omega^2} \
{1 \over v_0^2 p^2 + \omega^2} \cr
&\approx - {U_0 \over v_0^2} \left [ d_3 q^{d-4} {\omega^2 \over v_0^2}
+ d_4 q^{d-6} {\omega^4 \over v_0^4} \right ] \ , }
$$
$$
\eqalign{
d_3 &= {1 \over (4\pi)^{d/2}} \
{\Gamma (2-d/2) \ \Gamma^2 (d/2-1) \over \Gamma (d-2)} \ , \cr
d_4 & = {1 \over (4\pi)^{d/2}} \
{\Gamma (2-d/2) \ \Gamma (d/2-2) \ \Gamma (d/2-1) \over \Gamma (d-4)} \ , }
\eqno{(35)}
$$
where $v_0 q \gg \omega$ is assumed.
The bubble $\Pi_3$ of Fig.6 (b) is given by
$$
\eqalign{
\Pi_3 (q, \omega=0) &= - U_0
\int {d^d k \over (2\pi)^d} \ {d^d p \over (2\pi)^d} \ {d \nu \over 2\pi} \
\nu^2 \ {1 \over v_0^2 (k-q)^2 + \nu^2} \ {1 \over v_0^2 k^2 + \nu^2} \cr
&\times {1 \over v_0^2 (k-p)^2 + \nu^2}
\ {1 \over v_0^2 (k-p-q)^2 + \nu^2} \cr
&= - {U_0 \over v_0^5} \ f \
q^{d-3} q^{d-2} \ ,
}
\eqno{(36)}
$$
where
$$
f = {1 \over 2} \ {\Gamma (5/2 - d) \over (4 \pi)^{d + 1/2}}
\ \int_0^1 dx \int_0^1 dy \int_0^1 dz
{z^{1-d/2} (1-z)^{1-d/2} \over [ x (1-x) z + y (1-y) (1-z) ]^{5/2 - d} }
\eqno{(37)}
$$
is a convergent integral.
We can see that the bubble $\Pi_3$ is smaller by the factor $\epsilon$
( putting $d = 2 + \epsilon$ ) than $\Pi_1$ and $\Pi_2$ so that
the contribution from $\Gamma^{(4)}_{\rm bare,9}$ is higher order
in $\epsilon$ and we can neglect it.

Using the calculated $V_2$, we can evaluate $\Gamma^{(4)}_{\rm
bare}$ as
$$
\eqalign{
\Gamma^{(4)}_{\rm bare}
&= \sum_{i=1}^{9} \Gamma^{(4)}_{{\rm bare},i} \cr
&\approx \left ( {1 \over 1 + g_0 ( {\tilde \Pi}_1
+ \Pi_2 + \Pi_3 ) } \right )^2 W_0
+ 2 ( V_1 + V_2 ) \left ( {1 \over 1 + g_0 {\tilde \Pi}_1 }
\right )^2 W_0 \cr
&- U_0 \omega^2 ( 1 + 2 V_1 + 2 V_2 ) \cr
&\approx \left ( {1 \over g_0 ( {\tilde \Pi}_1 + \Pi_2 ) } \right )^2 W_0
+ 2 ( V_1 + V_2 ) \left ( {1 \over g_0 {\tilde \Pi}_1 } \right )^2 W_0
- U_0 \omega^2 ( 1 + 2 V_1 + 2 V_2 ) \cr
&\equiv \Gamma^{(4)}_{{\rm bare},w} + \Gamma^{(4)}_{{\rm bare},u} \ ,
}
\eqno{(38)}
$$
where
$$
\eqalign{
\Gamma^{(4)}_{{\rm bare},w}
&= {W_0 v^6 \over g_0^2 c_1^2} q^{2(3-d)} Z^{-4}
\left [ \left ( 1 + {W_0 \over v_0 g_0^2}{e \over c_1^3}
{v^3 \over Z^2} q^{2-d} \right )^{-2} +
2 {W_0 v_0^2 \over g_0^2} {d_1 \over c_1^2}
q^{2-d} \right. \cr
&- 2{U_0 \over v_0^2} \left. \left ( d_3 q^{d-4} {\omega^2 \over v_0^2}
+ d_4 q^{d-6} {\omega^4 \over v_0^4} \right ) \right ]  \ , \cr
\Gamma^{(4)}_{{\rm bare},u}
&= - U_0 \omega^2 \left [ 1 - 2 {U_0 \over v_0^2}
\left ( d_3 q^{d-4} {\omega^2 \over v_0^2}
+ d_4 q^{d-6} {\omega^4 \over v_0^4} \right ) \right. \cr
&+ 2 {W_0 v_0^2 \over g_0^2} \left. {d_1 \over c_1^2}
q^{2-d} \right ] \ , } \eqno{(39)}
$$
where $d_2 = 0$ for the short range interaction is used.

We take the follwoing renormalization condition for $\Gamma^{(4)}
= \Gamma^{(4)}_w + \Gamma^{(4)}_u$
$$
\eqalign{
\Gamma^{(4)}_{w} \bigr |_{q_1=q_4=\mu, \ q_2=q_3=0, \
\lim_{\alpha \rightarrow 0} \omega_1=\omega_2=\omega_3=\omega_4=\alpha v \mu}
&= {W v^6 \over g_0^2 c_1^2}\mu^{2(3-d)} \ , \cr
{\partial \Gamma^{(4)}_{u} \over \partial \omega^2}
\Bigr |_{q_1=q_4=\mu, \ q_2=q_3=0, \
\lim_{\alpha \rightarrow 0} \omega_1=\omega_2=\omega_3=\omega_4= \alpha v \mu}
&= \ - \ U \ . }
\eqno{(40)}
$$
Adding appropriate counter terms and using the renomalization
condition, we obtain the follwoing equations
$$
\eqalign{
{\tilde W_0}
&\approx {\tilde W} Z^2 \left ( {v_0 \over v} \right )^2
\left ( {\mu \over \Lambda} \right )^{d-2}
\left [ \ 1 + 2 {{\tilde W} \over c_1^2}
\ \left ( {{\tilde e} \over c_1} - {\tilde d_1} \right )
\ {\rm ln} \ {\Lambda \over \mu}
\ \right ] \ \cr
&\approx
{\tilde W}
\left ( {\mu \over \Lambda} \right )^{d-2}
\left [ \ 1 + {{\tilde W} \over c_1^2}
\ ( {\tilde c_3} + {\tilde c_2} + 2 {{\tilde e} \over c_1}
- 2 {\tilde d_1} ) \ {\rm ln} \ {\Lambda \over \mu}
- {\tilde U} \ {\tilde c_4}
\ {\rm ln} \ {\Lambda \over \mu}
\ \right ] \ , \cr
{\tilde U_0}
&\approx {\tilde U} Z^{-2} \left ( {v \over v_0} \right )^2
\left ( {\mu \over \Lambda} \right )^{2-d}
\left [ \ 1 - 2 {\tilde W} {{\tilde d_1} \over c_1^2}
\ {\rm ln} \ {\Lambda \over \mu}
\ \right ] \ \cr
&\approx
{\tilde U}
\left ( {\mu \over \Lambda} \right )^{2-d}
\left [ \ 1 - {{\tilde W} \over c_1^2} \ \left (
{\tilde c_3} + {\tilde c_2} + 2 {\tilde d_1} \ \right )
\ {\rm ln} \ {\Lambda \over \mu}
+ {\tilde U} \ {\tilde c_4}
\ {\rm ln} \ {\Lambda \over \mu}
\ \right ] \ .
}
\eqno{(41)}
$$
Using $\mu$ independence of the bare parameters, we obtain
the following equations for ${\tilde W}$ and ${\tilde U}$
$$
\eqalign{ \mu {\partial \over \partial \mu} {\tilde W} &\equiv
\beta ({\tilde W}) = - \epsilon {\tilde W} + {1 \over c_1^2}
\left [ {\tilde c_2} + {\tilde c_3} + 2 {{\tilde e} \over c_1}
- 2 {\tilde d_1} \right ] {\tilde W}^2
- {\tilde c_4} \ {\tilde U} {\tilde W} \ , \cr
\mu {\partial \over \partial \mu} {\tilde U} &\equiv
\beta ({\tilde U}) = \epsilon {\tilde U} - {1 \over c_1^2}
( {\tilde c}_3 + {\tilde c}_2 + 2 {\tilde d}_1 ) {\tilde W} {\tilde U}
+ {\tilde c}_4 \ {\tilde U}^2
\ , }
\eqno{(42)}
$$
where ${\tilde c}_4 = \epsilon c_4$ and .
For small $\epsilon$, the coefficients of these equations are given by
$$
c_1 = 1/8, \ {\tilde c}_2 = 1/\pi, \ {\tilde c}_3 = - 1/\pi, \
{\tilde c}_4 = 1/(2\pi), \ {\tilde d}_1 = 1/\pi \ .
\eqno{(43)}
$$
We can ignore ${\tilde e}$ which is the order of $\epsilon$.
Now the RG equations for ${\tilde W}, {\tilde U}$ and $v$ are given by
$$
\eqalign{ {d {\tilde W} \over d l} &= - \beta ({\tilde W}) =
\epsilon {\tilde W} + {1 \over 2 \pi} {\tilde U} {\tilde W} +
{128 \over \pi} {\tilde W}^2  \ , \cr
{d {\tilde U} \over d l} &= - \beta ({\tilde U}) =
- \epsilon {\tilde U} + {128 \over \pi} {\tilde W} {\tilde U}
- {1 \over 2 \pi} {\tilde U}^2  \ , \cr
{d v \over d l} &= - \beta (v) =  - {64 \over \pi} v {\tilde W}
- {1 \over 2\pi} v {\tilde U} \ ,}
\eqno{(44)}
$$
where $l$ is agian the logarithmic measure of the RG flow.
The RG flow is drawn in Fig.7 (a) for $\epsilon \ge 0$, (b) for
$\epsilon < 0$.

For $d < 2 \ ( \epsilon < 0 )$, we can see that there are three fixed
points which are given by $({\tilde W}^*, {\tilde U}^* ) = (0, 0),
\ (0, 2\pi |\epsilon|),
\ ({\pi \over 128} |\epsilon|, 0)$.
The dynamical exponents at these fixed points are given by
$z = 1, \ 1 + |\epsilon|, \ 1 + {1 \over 2} |\epsilon|$ respectively.
Note that all of these fixed points are essentially unstable because
they have at least one relevant direction in the ${\tilde W}-{\tilde U}$
plane.
Among these three fixed points, $({\tilde W}^*, {\tilde U}^* )
= (0, 2\pi |\epsilon|)$ is a non-trivial fixed point that has the least
number (one) of relevant directions. However, we need to fine tune the
strength of the $w({\bf x})$ type disorder in order to reach this fixed point.
For $2 \le d < 3 \ ( \epsilon \ge 0 )$, there is only the trivial fixed point
$({\tilde W}^*, {\tilde U}^* ) = (0, 0)$. At this fixed point,
${\tilde U}$ is relevant and ${\tilde W}$ is irrelevant so that the RG
flow goes to the strong randomness regime where our RG scheme breaks down.

Therefore, in both cases, there is no stable non-trivial random fixed
point near the pure $XY$ fixed point.
This means that the SF-insulator transition should be discribed
by possible strong random fixed points. The conclusion we can deduce
from these results is that, near $d=2$, the direct SF-MI transition is
unlikely to happen in the whole ${\bar w}-{\bar u}$ plane
even for the weak disorder. This result essentially
support the original picture of Fisher {\it et al.}[1,2] (Fig.1 (a))
that the SF-insulator transition should alwalys occur from the Bose glass
phase rather than from the Mott insulator even for the
commensurate case.

\vskip 0.5cm

\centerline{\bf IV. 1/N CORRECTION}

In this section, we are going to investigate the effects of the $1/N$
correction to the RG equation.
The $1/N$ correction due to the $\phi^4$ interaction
in the previous commensurate dirty boson model will be considered.
The most important $1/N$ correction that can affect the
RG equation enters in the coefficient of the ${\tilde W}$ term in
Eq.(44).
This $1/N$ correction can be read off from the scaling dimensions
of $\phi^{\dagger} \phi$ at the pure fixed point [20]:
$$
[ \ \phi^{\dagger} \phi \ ] = 2 - \eta \ ,
\eqno{(45)}
$$
where
$$
\eta = {32 \over 3 \pi^2} {1 \over 2N} \ .
\eqno{(46)}
$$
As a result, the critical dimension for ${\tilde W}$ is changed from $2$ to
$2 - 2 \eta$. If the convention of $\epsilon = d - 2$ was taken,
the RG equation becomes
$$
\eqalign{ {d {\tilde W} \over d l} &=
( \epsilon + 2 \eta ) {\tilde W} + {1 \over 2 \pi} {\tilde U} {\tilde W} +
{128 \over \pi} {\tilde W}^2  \ , \cr
{d {\tilde U} \over d l} &=
- \epsilon {\tilde U} + {128 \over \pi} {\tilde W} {\tilde U}
- {1 \over 2 \pi} {\tilde U}^2  \ , \cr
{d v \over d l} &= - {64 \over \pi} v {\tilde W}
- {1 \over 2\pi} v {\tilde U} \ .}
\eqno{(47)}
$$
Note that the density $\rho$ has no anormalous dimension [16] and the RG
equation for ${\tilde U}$ is not affected at this order.

Now there can be three possible RG flows that depend on the dimensionality
of the system and $N$.
For $\epsilon = d - 2 \ge 0$, there is only one fixed point which is the
unstable trivial fixed point $({\tilde W}^*, {\tilde U}^*) = (0,0)$. The RG
flow is given by Fig.7 (a) and it flows to the strong disorder regime.
For $\epsilon < 0$ and $|\epsilon| < 2\eta$, there can be two fixed points
which are $({\tilde W}^*, {\tilde U}^*) = (0,0), \ (0,2 \pi |\epsilon|)$ with
the dynamical exponents $z = 1, \ 1 + |\epsilon|$. But all of them are
unstable and the RG flow is given by Fig.8.
On the other hand, for $\epsilon < 0$ and $|\epsilon| > 2\eta$, there is
one more fixed point which is given by $({\tilde W}^*, {\tilde U}^*) =
({\pi \over 128}(|\epsilon|-2\eta),0)$ with
$z = 1 + {1 \over 2} (|\epsilon|-2\eta)$. All of the three fixed points
are still unstable and the RG flow goes to the strong disorder regime. The
RG flow is again given by Fig.7 (b).
Therefore, after the inclusion of the $1/N$ correction, there is no
stable weak random fixed point.
The direct SF-MI transition is unlikely in the whole
${\bar w}-{\bar u}$ plane.

\vskip 0.5cm

\centerline{\bf V. SUMMARY AND CONCLUSION}

We study a large N generalization of the commensurate dirty boson problem.
The ${1 \over N}$ expansion allows us to treat interaction
effects properly.
On the other hand, the
disorder is assumed to be weak and the pertubation in the strength of
the randomness is performed.
In order to understand the behaviors of bosons in this model, we
need two types of disorder, {\it i.e.}, the random coefficient
of the density term $\phi^{\dagger} \ \partial_0 \phi$, $u({\bf x})$,
and the random coeffient of the quadratic term
$\phi^{\dagger} \phi$, $w({\bf x})$.

For a restricted model with $u({\bf x})=0$ which has an additional
particle-hole symmetry, we introduce more general
long range interactions $V(q) = g_0 / q^{\lambda}, ( 0 \le \lambda < 1/2 )$
( which is not a genuine long range interaction as mentioned in Sec. II ).
The critical dimension for $w({\bf x})$ type disorder is found to be
$d_c = 2 + 2 \lambda$ and we performed $\epsilon = d - 2 - 2\lambda$
expansion.
It is found that $\lambda = 0$ case and $\lambda \not= 0$ case show different
behaviors and they are not continuously related in the $\epsilon = d - d_c$
expansion.

First, let us look at the case of $\lambda \not= 0$.
For $d > d_c$, the pure fixed point is unstable, but there is a stable
fixed point which governs the transition. For $d \le d_c$, the pure
fixed point is stable and the disorder is irrelevant.
Therefore, we expect the direct SF-MI transition.
For $d \le d_c$, the critical point is the same as that for the pure system.
For $d > d_c$, SF-MI transition is described by a new non-trivial fixed
point.

For the short range interaction ($\lambda = 0$), the disorder is
relevant for $d \ge 2$ and irrelevant for $d < 2$.
There is only the unstable pure fixed point and the RG flow goes to the
strong disorder regime for $2 \le d < 3$. For $d < 2$, the pure fixed
point becomes stable and there is also an unstable non-trivial fixed point.
Therefore, we expect a direct SF-MI transition for $d < 2$ but
none for $2 \le d < 3$.

We consider the usual short range interaction for the general case
of $u({\bf x}) \not = 0$ but ${\bar u} = 0$
(the commensurate dirty boson problem).
The critical dimension of both types of disorder is found to be
$d = 2$ (at $(1/N)^0$ order) and we performed $\epsilon = d - 2$
expansion ($d < 3$).
For $d < 2 $, we have three fixed points and they have at least one
relevant direction in the ${\tilde W}-{\tilde U}$ plane.
In the case of $2 \le d < 3$, the RG flow is
governed by the $w({\bf x})$ type disorder and it flows to the
strong disorder regime which cannot be reached by the pertubation in
the strength of the randomness.
Therefore, we expect that the direct SF-MI transition is unlikely to
happen near two dimensions.

The effects of the $1/N$ correction are considered.
For the commensurate dirty boson model, the $1/N$ correction due to the
$\phi^4$ interaction does not change the qualitative features of the
problem. There is still no stable weak random fixed point for $d \sim 2$
and the direct SF-MI transition is unlikely.

\vskip 0.5cm

\centerline{\bf ACKNOWLEDGMENTS}

We thank P. A. Lee, M. Kardar, C. Chamon for helpful discussions. This work
is supported by NSF grant No. DMR-91-14553.

\vfill\vfill
\break

\vskip 0.5cm

\centerline{\bf REFERENCES}

\item{[1]} D. S. Fisher and M. P. A. Fisher, Phys. Rev. Lett. {\bf 61},
1847 (1988)
\item{[2]} M. P. A. Fisher, P. B. Weichmann, G. Grinstein,
and D. S. Fisher, Phys. Rev. B {\bf 40}, 546 (1989)
\item{[3]} E. S. Sorensen, M. Wallin, S. M. Girvin, and A. P. Young,
Phys. Rev. Lett. {\bf 69}, 828 (1992)
\item{[4]} M. Ma, B. I. Halperin, and P. A. Lee, Phys. Rev. B {\bf 34},
3136 (1986)
\item{[5]} K. G. Singh and D. S. Rokhsar, Phys. Rev. B {\bf 46},
3002 (1992)
\item{[6]} L. Zhang and M. Ma, Phys. Rev. B {\bf 45},
4855 (1992)
\item{[7]} P. B. Weichman and K. Kim, Phys. Rev. B {\bf 40},
813 (1989)
\item{[8]} W. Krauth, N. Trivedi and D. Ceperley, Phys. Rev. Lett. {\bf 67},
2307 (1991)
\item{[9]} M. Makivic, N. Trivedi and S. Ullah, Preprint.
\item{[10]} D. K. K. Lee and J. M. F. Gunn, J. Phys. Condens. Matter {\bf 2},
7753 (1990)
\item{[11]} P. A. Lee and T. V. Ramakrishnan, Rev. Mod. Phys. {\bf 57},
287 (1985)
\item{[12]} J. D. Reppy, Physica {\bf 126 B},
335 (1984)
\item{[13]} M. H. W. Chan, K. I. Blum, S. Q. Murphy, G. K. S. Wong,
and J. D. Reppy, Phys. Rev. Lett. {\bf 61},
1950 (1988)
\item{[14]} H. M. Jaeger, D. B. Haviland, B. G. Orr, and
A. M. Goldman, Phys. Rev. B {\bf 40},
182 (1989)
\item{[15]} M. Randeria, J. M. Duan and L. Y. Shieh,
Phys. Rev. Lett. {\bf 62}, 981 (1989)
\item{[16]} X. -G. Wen and Y. -S. Wu, Phys. Rev. Lett. {\bf 70},
1501 (1993)
\item{[17]} W. Chen, M. P. A. Fisher, and Y. -S. Wu, Preprint
\item{[18]} J. A. Hertz, L. Fleishman and P. W. Anderson,
Phys. Rev. Lett. {\bf 43}, 942 (1979)
\item{[19]} E. Brezin, J. C. Le Guillon, and J. Zinn-Justin,
\hskip 0.2cm {\it Phase transitions and critical phenamena},
Vol 6, C. Domb and M. S. Green eds. ( Academic Press, New York, 1976 )
\item{[20]} S. K. Ma, {\it Modern theory of critical phenamena},
( Addison-Wesley, Reading, MA, 1976 )

\vfill\vfill
\break

\vskip 0.5cm

\centerline{\bf FIGURE CAPTIONS}

\item{Fig.1}
Three possible phase diagrams for the dirty boson system described by
Eqs. (2) and (3). Our RG results favor the case (a).
The boson density commensurates with the lattice in the MI phase and
on the dotted line.
In (a) there is no direct SF-MI transition.
In (b) direct SF-MI transition is described by a tricritical point.
Both of the BG and the MI phases are insulators, but the former
is gapless and the latter has a finite gap.
The commensurate dirty boson model is defined on the dotted line
and its extension in the MI phase.

\hskip 0.5cm

\item{Fig.2}
(a) Self energy correction due to the $w({\bf x})$ type disorder.
The wavy line represents the interaction and the dashed line with a cross
$\times$ means the impurity average. $W$ means $w({\bf x})$ type disorder.
(b) The polarization bubble $\Pi_1$.
(c) The renormalized interaction ( thick line )
due to the $w({\bf x})$ type disorder.

\hskip 0.5cm

\item{Fig.3} (a), (b), (c), (d) are diagrams that contribute to
the four point function up to ${\tilde W}^2$ order.
(e) The renormalzed polariztion bubble ${\tilde \Pi}_1$. The double
line means the renormalized full propagator.
(f) The polariztion bubble $\Pi_2$.
(g) The vertex $V_1$.

\hskip 0.5cm

\item{Fig.4} The self energy correction due to the $u({\bf x})$ type
disorder. The dashed line with a cross $\times$ means the impurity
average. $U$ means the $u({\bf x})$ type disorder.

\hskip 0.5cm

\item{Fig.5} Additional diagrams that contribute to the four point
function in the generic commensurate dirty boson model up to 2nd order
in the impurity stregth.

\hskip 0.5cm

\item{Fig.6} (a) The vertex $V_2$. (b) The polarization bubble $\Pi_3$.

\hskip 0.5cm

\item{Fig.7} RG flows (a) for $2 \le d < 3$ and (b) for $d < 2$.

\hskip 0.5cm

\item{Fig.8} RG flows which has the $1/N$ correction for $d < 2$ and
$|\epsilon| < 2\eta $.

\vfill\vfill

\bye